\documentclass[a4paper,11pt]{article}
\usepackage{jheppub}
\usepackage{lineno}

\usepackage{blindtext}
\usepackage{mathtools}
\usepackage{xcolor}

\usepackage{mathrsfs}
\usepackage{braket}

\usepackage[utf8]{inputenc}
\usepackage[english]{babel}
\usepackage{hyperref}
\definecolor{mypink1}{rgb}{0.858, 0.188, 0.478}
\definecolor{mypink2}{RGB}{219, 48, 122}
\hypersetup{
    colorlinks=true,
    linkcolor=blue,
    filecolor=blue,      
    urlcolor=blue,
    citecolor=blue,
}
\usepackage{color,colortbl}
\definecolor{LightCyan}{rgb}{0.88,1,1}
\usepackage{tensind}
\tensordelimiter{?}
\newcommand{\dd}{\!{\mathrm{d}}}

\usepackage{graphicx}
\usepackage{hyperref}
\DeclareGraphicsExtensions{.bmp,.png,.jpg,.pdf}
\usepackage{amsmath}
\usepackage{courier}
\usepackage{verbatim}
\usepackage{amssymb}
\usepackage{amsfonts}
\usepackage{natbib}
\usepackage{soul}

\usepackage{dsfont}

\usepackage{booktabs}
\usepackage{siunitx}
\usepackage{cleveref}

\newcommand{\Btrap}{\mathbf{B}}
\newcommand{\Bdm}{\mathbf{B}_\mathrm{DM}}
\newcommand{\Bloop}{\mathbf{B}_\mathrm{loop}}

\newcommand{\Jeff}{\mathbf{J}_\mathrm{eff}}

\newcommand{\vr}{\mathbf{r}}
\newcommand{\mfp}{m_{\mathrm{FP}}}

\usepackage{yfonts}
\newcommand{\alphaM}{\alpha_\mathfrak{m}}
\newcommand{\alphaL}{\alpha_\mathfrak{l}}

\DeclareSIUnit{\yr}{yr}

\newcommand{\NTE}[1]{\mathcal{N}^{#1}_{\mathrm{TE}}}
\newcommand{\NTM}[1]{\mathcal{N}^{#1}_{\mathrm{TM}}}

\arxivnumber{2603.22647}
\title{Dark graviton sensing with magnetically levitated superconductors}
\author{Valentina Danieli, Paola C.~M.~Delgado and Federico R.~Urban}
\affiliation{CEICO, FZU – Institute of Physics of the Czech Academy of Sciences\\ Na Slovance 1999/2, 182 00 Prague, Czech Republic}
\emailAdd{danieli@fzu.cz}
\emailAdd{delgado@fzu.cz}
\emailAdd{federico.urban@fzu.cz}

\abstract{Levitated sensors have emerged as a new frontier to detect ultra-light dark matter such as axion-like particles and dark photons. In this work we study how a magnetically levitated superconductor responds to a spin-2 dark matter field, the dark graviton, in the dHz to kHz frequency range. To do so, we compute the forces that the dark graviton exerts on the superconductor, separately for matter and light couplings. The matter coupling produces a strain-like tidal acceleration between the superconductor and the centre of mass of the rigid apparatus, resulting in a displacement relative to the readout pick-up loop, in a way analogous to the response to a slow, continuous, massive gravitational wave. The light coupling instead induces an effective current that sources an oscillating magnetic field, thus driving the superdiamagnetic response of the superconductor. We find that, albeit with significant experimental improvements, the sensitivity reach for the matter coupling could surpass existing fifth-force experiments in the audio frequency band. At the same time, magnetically levitated superconductors could be among the most sensitive laboratory probes of the dark-graviton coupling to electromagnetism, especially at low frequencies, provided technical and readout noise can be kept under control.}

\begin{document}
\maketitle
\flushbottom

\section{Introduction}

The search for dark matter (DM), whose existence we infer from cosmological and astrophysical observations, is among the most active areas of astroparticle and particle physics research today~\cite{Bozorgnia:2024pwk}. Astrophysical and laboratory tests, on Earth or in space, aim to cover several dozen orders of magnitude for the unknown DM mass, as well as a profusion of theorised couplings between DM and standard matter. One of the most compelling classes of dark matter candidates is ultra-light dark matter (ULDM): weakly interacting slim particles~\cite{Albertus:2026fbe} with a mass well below the~eV~\cite{Ferreira:2020fam}. Owing to its lightness, in order to account for the whole cosmological dark matter, ULDM must have integer spin; hence, since this also implies a very large occupation number, ULDM can be approximated by a superposition of classical waves~\cite{Centers:2019dyn}. Driven by both tremendous experimental advances in quantum sensing and theoretical developments, ULDM has flourished as a fertile and versatile family of models that can address unresolved issues in astroparticle and particle physics~\cite{Arza:2026rsl}.

Within the vast landscape of experimental approaches that have been developed to test ULDM, levitated sensor detectors (LSDs)~\cite{Gonzalez-Ballestero:2021gnu} have proven to be remarkably functional and adaptable in order to probe the interactions between ULDM and standard matter~\cite{Carney:2019cio,Carney:2020xol,Moore:2020awi,Kilian:2024fsg}. In an LSD, a test mass is levitated in high vacuum by means of optical, electrical or magnetic traps, and its motion is carefully monitored with precision readout systems. The ability to control the motion of the sensor and the accuracy of the, often quantum-limited, readouts allows the detection of minute forces acting on the sensor~\cite{Carney:2019cio,Carney:2020xol,Moore:2020awi,Kilian:2024fsg}.

Among LSD architectures, magnetically levitated sensors are attractive experimentally because they avoid optical absorption and photon-recoil heating, can operate cryogenically and naturally interface with superconducting circuitry for low-noise readout~\cite{Carney:2020xol,Latorre:2022vxo,Hofer:2022chf}. From a theoretical perspective, magnetic levitation is attractive because it is sensitive to both inertial forces (probing couplings to matter) and electromagnetic forces, opening discovery channels for dark matter couplings to photons. Magnetically levitated sensors, acting as accelerometers and/or magnetometers, have been proposed to detect axions, axion-like particles and vector bosons~\cite{Graham:2015ifn,Manley:2020mjq,Windchime:2022whs,Li:2023wcb}.

As a promising example of such technology, in~\cite{Higgins:2023gwq} it was proposed to use a superconducting particle (SCP) levitated by means of a magnetic trap. In this set-up, the interaction between the ULDM and electromagnetism provides a coherently oscillating perturbation in the surface currents that keep the SCP at its equilibrium point, thereby driving the motion of the SCP. This generates a variable magnetic flux that is picked up by the readout system, which translates into a measurement of the interaction strengths between the ULDM and photons.\footnote{A variation on this idea, which has been already demonstrated experimentally, consists of levitating a permanent magnet over a superconducting plane~\cite{Amaral:2024rbj}; yet another option is to use a ferromagnet, see~\cite{Vinante:2022hnf} for a review and~\cite{Kalia:2024eml} for its application to ULDM detection.} This technique has the potential to be highly competitive with other laboratory ULDM searches, in particular for the dark photon DM case~\cite{Higgins:2023gwq}.

In this work we extend the analysis of~\cite{Higgins:2023gwq} to the case of spin-2 ULDM, the dark graviton (also known as the dark tensor)~\cite{Aoki:2016zgp,Babichev:2016hir,Babichev:2016bxi,Aoki:2017cnz,Marzola:2017lbt}. We derive the induced matter and electromagnetic forces on a magnetically levitated SCP sourced by a non‑relativistic dark graviton and show expected sensitivity projections in various set-ups, for a dark graviton mass spanning \(10^{-16}\,\mathrm{eV}\lesssim\mfp\lesssim10^{-11}\,\mathrm{eV}\). To this end, we exploit the LSD's concomitant sensitivity to tidal (matter) forces as well as electromagnetic (light) forces. This is particularly germane to this context because the same LSD technology underpins some of the proposals for compact, high-frequency gravitational-wave detectors~\cite{Carney:2024zzk} and, from the point of view of the LSDs, the dark graviton acts like a slow, massive, continuous gravitational wave, producing a strain-like monochromatic signal.

This paper is organised as follows. In the next \cref{sec:set-up} we review the basics of the experimental set-up. Then, in \cref{sec:spin-2} we introduce our formalism for the dark graviton and derive the forces experienced by the SCP caused by its coupling to matter and to light. We present our sensitivity forecasts in \cref{sec:sensitivity} alongside a discussion of our results. We conclude and provide an outlook for future work in \cref{sec:conclusion}.

\section{Set-up}
\label{sec:set-up}

To begin with, let us review the experimental set-up proposed in~\cite{Higgins:2023gwq}. The experimental apparatus consists of a SCP that is suspended against the gravitational pull by a magnetic trap. The trap is composed of two circular coils with currents flowing in opposite directions (anti-Helmholtz-like configuration), whose generated magnetic field gradients keep the SCP at the centre of the trap. The trap-SCP system is enclosed in a magnetic shield whose main purpose is to isolate the system from external fields. Changes in the magnetic field generated by perturbations in the equilibrium position of the SCP are picked up by a readout system such as a superconducting quantum interference device (SQUID), typically placed above the SCP.

The trap magnetic field in this configuration can be expressed, at linear order around the trap centre as
\begin{align}
\label{eq:Bi}
    B_i (\mathbf{x},t)=B_{0,i}(t)+b_{ij}(t)x_j,
\end{align}
where $B_0$ denotes the magnetic field at the centre of the trap and $b_{ij}$ represents the field gradients in its proximity. In isolation, i.e.~when the dark graviton is not present, the only contribution to the magnetic field within the shield comes from the trapping quadrupolar field, which is static. Therefore, it is always possible to find a coordinate system in which $B_{0_i}(t)=0$ and $b_{ij}$ is diagonal. Consequently, the trapping field can be written as
\begin{align}
\label{eq:Btrap}
    \Btrap(\mathbf{x},t)=b_{xx}x\hat{x}+b_{yy}y\hat{y}+b_{zz}z\hat{z}.
\end{align}

Let us now consider a SCP with volume $V_\mathrm{SCP}$, positioned at \(\mathrm{r}_0\) within the magnetic field $\Btrap$ given by \cref{eq:Btrap}. If the SCP does not sit at the centre of the trap, surface currents are generated on the surface of the SCP that eject the magnetic field out of its interior, effectively screening the SCP. Such surface currents, because of the trapping magnetic field, are subject to a Lorentz force which pulls the SCP back to the centre of the trap. The restoring force is given by
\begin{align}
\label{eq:Lorentzforce}
    F_i(\mathbf{x},t)=-\frac{3}{2}V_\mathrm{SCP} b_{ij} B_j(\mathbf{x},t),
\end{align}
see~\cite{Hofer_2019} for a derivation.\footnote{Notice that the net force experienced by the SCP consists of the sum of the forces acting on the whole surface of the SCP; this force consequently depends also on the gradients $b_{ij}$ of the magnetic field across the SCP.} Substituting the trapping field \cref{eq:Btrap} into \cref{eq:Lorentzforce}, the trapping force reads
\begin{equation}
\label{eq:Ftrap}
    \mathbf{F}(\mathbf{x},t) = -\frac{3}{2}V_\mathrm{SCP} \left( b^2_{xx}x\hat{x}+b^2_{yy}y\hat{y}+b^2_{zz}z\hat{z} \right),
\end{equation}
indicating that the system behaves as a harmonic oscillator. The corresponding resonant frequencies are
\begin{align}
    f_i\coloneqq\sqrt{\frac{3}{8\pi^2\rho}}b_{ii}
\end{align}
(no summation over \(i\)), with $\rho$ the density of the SCP.

If an external field interacts directly with the experimental apparatus in such a way that the equilibrium position of the SCP changes with time, it effectively acts as a driver for this harmonic oscillator. In the case of the dark graviton two effects have to be taken into account. Firstly, the direct coupling with matter generates a force that displaces the SCP from its equilibrium position, thus triggering the oscillatory motion. Secondly, the coupling with light generates an effective current $\Jeff$, sourcing an effective ``dark matter'' magnetic field $\Bdm$, which in turn gives rise to an effective Lorentz force acting on the SCP. In the following section we will consider both these effects.

\section{The dark graviton signal}
\label{sec:spin-2}

Let us consider a massive spin-2 particle (the dark graviton) described by the Lorentz-invariant, but not necessarily universally coupled, Fierz-Pauli Lagrangian
\begin{align}\label{eq:s2lag}
    \mathcal{L}_{\rm FP} \coloneqq -\varphi^{\mu\nu}{\mathcal E}_{\mu\nu}{}^{\rho\sigma}\varphi_{\rho\sigma}-\frac{1}{2}\mfp^2(\varphi_{\mu\nu}\varphi^{\mu\nu}-\varphi^2),
\end{align}
where $\varphi = \eta^{\mu\nu} \varphi_{\mu\nu}$ and the Lichnerowicz operator is defined as
\begin{align}
    {\mathcal E}_{\mu\nu}{}^{\rho\sigma}\varphi_{\rho\sigma} \coloneqq -\frac{1}{2}(\Box \varphi_{\mu\nu}-\partial_{\mu}\partial^\alpha \varphi_{\alpha\nu}-\partial_\nu \partial^\alpha \varphi_{\alpha\mu} +\partial_\mu \partial_\nu \varphi -\eta_{\mu\nu}\Box\varphi+\eta_{\mu\nu}\partial_\alpha \partial_\beta\varphi^{\alpha\beta}),
\end{align}
with $\Box \coloneqq \eta^{\mu\nu}\partial_\mu \partial_\nu$~\cite{deRham:2014zqa}. This phenomenological description can be thought of as the low-energy limit of a ghost-free non-linear completion in the EFT sense~\cite{Albertus:2026fbe}, which implies that the scales we can probe must be within the regime of validity of this EFT. The coupling of the dark graviton with the Standard Model can be generically written as
\begin{equation}
\label{eq:couplings}
    \mathcal{L}_\mathrm{int} = \varphi_{\mu\nu}\, \mathcal{O}^{\mu\nu},
\end{equation}
where $\mathcal{O}_{\mu\nu}$ is a symmetric tensor built out of Standard Model fields -- in Lorentz-invariant, universally coupled theories this is proportional to the energy-momentum tensor \(T^{\mu\nu}\). In the non-relativistic limit, which is the regime valid for our experiment, the possible couplings to matter and light are respectively given by
\begin{equation}
\label{eq:Lmatter}
    \varphi_{\mu\nu} \mathcal{O}^{\mu\nu}_\mathfrak{m}=\frac{\alphaM}{\Lambda} \varphi_{\mu\nu} \frac{P^\mu P^\nu}{M} \delta^{(3)}\left(\vr-\vr_0\right),
\end{equation}
and
\begin{align}
\label{eq:Llight}
    \varphi_{\mu\nu} \mathcal{O}^{\mu\nu}_\mathfrak{l} = \frac{\alphaL}{\Lambda}\varphi_{\mu\nu} F^\mu{}_\lambda F^{\nu\lambda}.
\end{align}
Here we have defined the dimensionless matter and light coupling constants \(\alphaM\) and \(\alphaL\), respectively, $M$ denotes the mass of the SCP (located at \(\vr=\vr_0\)), $P^\mu$ refers to its four-momentum $P^\mu = (M,\mathbf{P})$ and $\Lambda$ is the mass scale that characterises the interactions with the dark graviton.\footnote{Here by `light' we mean the electromagnetic sector of the Standard Model; in practice, as we will see, in this case the coupling is effectively to the magnetic field.} We have neglected possible couplings involving the trace of the dark graviton in both interaction terms because the equations of motion for the free \(\varphi_{\mu\nu}\) field imply its tracelessness and transversality, namely $\varphi^\mu{}_\mu =0 = \partial_\mu \varphi^{\mu\nu}$, rendering the effects of such terms subdominant~\cite{Marzola:2017lbt}. Lastly, even though \cref{eq:Lmatter} and \cref{eq:Llight} do not fully define the microscopic theory, all leading physical effects in our set-up are fully captured by these two interaction terms.

Because in our local environment the DM waves are very slow at \(v_{\rm DM}\sim10^{-3}\), the dark graviton is non-relativistic; its tracelessness and transversality enforce the gradient hierarchy $\varphi_{00} \sim v_{\rm DM}^2 \varphi_{ij}$ and $\varphi_{0i} \sim v_{\rm DM} \varphi_{ij}$, so that we can safely neglect $\varphi_{0\mu}$. Moreover, because the whole experiment is well within the DM de Broglie wavelength, \(\lambda_\mathrm{dB} \coloneqq 2\pi / (\mfp v_\mathrm{DM})\), we can ignore gradients throughout, which is equivalent to ignoring the dark graviton momentum \(\mathbf{k}\). Thus, the dynamics of the dominant DM field components is expressed as
\begin{align}
\label{eq:dmansatz}
    \varphi_{ij} = \varphi_0 \cos{(\mfp t +\zeta )}\, \varepsilon_{ij},
\end{align}
where \(\varphi_0 = \sqrt{2\rho_{\rm DM}}/\mfp\) is the field amplitude, determined by the requirement that \(\varphi_{ij}\) makes up all of the DM, $\rho_{\rm DM} = 0.4\,\mathrm{GeV}/\mathrm{cm}^3$ is the DM energy density~\cite{Read:2014qva}, $\zeta$ is a random phase and $\varepsilon_{ij}$ is the polarisation tensor that encodes the five physical degrees of freedom of the dark graviton.\footnote{In some of the literature the alternative normalisation \(\varphi_0 = \sqrt{2\rho_{\rm DM}}/\sqrt5\mfp\) is used to explicitly account for the fact that the field \(\varphi_{ij}\) propagates five degrees of freedom; we encode the \(1/\sqrt5\) factor in the normalisation of the polarisation tensor, \cref{eq:polT} below.} In the orthonormal DM basis $(\mathbf{p},\mathbf{q},\mathbf{k})$, the polarisation tensor is defined as 
\begin{equation}\label{eq:polT}
    \varepsilon_{ij}(\mathbf{k})\coloneqq\sum_{\kappa} \varepsilon_{\kappa} \mathcal{Y}_{i j}^{\kappa}(\mathbf{k}),
\end{equation}
where the summation runs over the five amplitudes $\{\varepsilon_\times,\,\varepsilon_+,\,\varepsilon_L,\,\varepsilon_R\,,\varepsilon_S\}$, satisfying $\sum_\kappa \varepsilon_\kappa^2 =1$ and
\begin{equation}\label{eq:pols}
\begin{aligned}
    &\mathcal{Y}_{i j}^{\times}\coloneqq\frac{1}{\sqrt{2}}\left(p_{i} q_{j}+q_{i} p_{j}\right), \quad & \mathcal{Y}_{i j}^{+}\coloneqq\frac{1}{\sqrt{2}}\left(p_{i} p_{j}-q_{i} q_{j}\right), \\
    & \mathcal{Y}_{i j}^{L}\coloneqq\frac{1}{\sqrt{2}}\left(q_{i} k_{j}+k_{i} q_{j}\right), \quad & \mathcal{Y}_{i j}^{R}\coloneqq\frac{1}{\sqrt{2}}\left(p_{i} k_{j}+k_{i} p_{j}\right), \\
    & \mathcal{Y}_{i j}^{S}\coloneqq\frac{1}{\sqrt{6}}\left(3 k_{i} k_{j}-\delta_{i j}\right), &
\end{aligned}
\end{equation}
see for instance~\cite{Armaleo:2020efr}. The DM basis is linked to the proper detector frame \((\mathbf{x},\mathbf{y},\mathbf{z})\) via a \((\theta,\phi)\) rotation
\begin{equation}
\begin{aligned}
    \mathbf{k}& \coloneqq (\sin \theta \cos \phi, \sin \theta \sin \phi, \cos \theta),\\
    \mathbf{p}& \coloneqq (\cos \theta \cos \phi, \cos \theta \sin \phi,-\sin \theta),\\
    \mathbf{q}& \coloneqq (-\sin \phi, \cos \phi, 0).
\end{aligned}
\end{equation}

The local DM environment is in fact composed of a large superposition of DM waves, each with its own velocity distributed according to a Maxwell-Boltzmann distribution centred around the virial velocity of DM, its own polarisation and phase~\cite{Centers:2019dyn}; however for the order-of-magnitude estimates we present in this work we will ignore this except when folding it into our sensitivity projections in \cref{sec:sensitivity}. We now proceed to consider the effects of the two couplings \cref{eq:Lmatter} and \cref{eq:Llight} in turn, assuming that each is parametrically larger than the other in their respective sections, and discuss the universal case, \(\alphaM=\alphaL\) separately in \cref{sec:sensitivity}. Lastly, here and in what follows, the proper detector frame is operationally defined as the frame in which all matter elements of the detector respond to a locally flat metric; this frame corresponds to the usual proper detector frame in the universally coupled limit.

\subsection{Matter coupling}
\label{sec:matter}

The effect of the dark graviton on the experimental apparatus is akin to that of a slow, massive, continuous gravitational wave, and we can thus employ the same formalism~\cite{Maggiore:2007ulw}. This can be seen explicitly by noticing that, at the linear level, because the interaction \cref{eq:Lmatter} corresponds to a linear coupling to the energy-momentum tensor defined as the variation of the matter action with respect to the metric, the matter coupling can be reabsorbed into a redefinition of the metric itself as
\begin{equation}\label{eq:metric_re}
    \tilde{g}_{\mu\nu} = g_{\mu\nu} + \frac{2\alphaM}{\Lambda} \varphi_{\mu\nu}.
\end{equation}
The effect of such a field in the proper detector frame, where the metric is locally flat along the detector worldline, is captured by the geodesic deviation equation
\begin{align}\label{eq:geodev}
    \frac{\dd^2 \xi^i}{\dd\,t^2} \coloneqq - R^i_{\;0j0} \xi^j = \frac{\alphaM}{\Lambda}\left( \ddot\varphi_{ij} - \partial_i \dot\varphi_{0j} - \partial_j \dot\varphi_{0i} + \partial_i \partial_j \varphi_{00} \right)\xi^j \simeq \frac{\alphaM}{\Lambda} \ddot\varphi_{ij}\xi^j,
\end{align}
where, unlike for a gravitational wave, the gradient terms are not zero by gauge choice (for instance, the transverse-traceless gauge) but because of the gradient hierarchy imposed by the tracelessness and transversality conditions that descend from the dark graviton equations of motion in vacuum.\footnote{Notice that the factor of \(2\) in \cref{eq:metric_re} arises from the canonical \(1/2\) normalisation of the dark graviton in \cref{eq:s2lag} compared to the \(1/8\) normalisation common in General Relativity~\cite{deRham:2014zqa}.} Note that, because the mechanical vibrations caused by the dark graviton on the experimental apparatus (shield, SCP, coils and pick-up loop) travel at a sound speed of about \(c_s \approx 10^{-5}\), for DM frequencies below \(1\,\mathrm{kHz}\) the whole experimental apparatus can be considered as rigid. Lastly, as previously mentioned, we can ignore gradients and all inertial forces that act on the SCP, as they are suppressed by a factor \(L / \lambda_\mathrm{dB} \ll1\) with \(L\) being the length of the (cubic) shield.

The dark graviton causes a relative displacement between the SCP and the rest of the apparatus, and in particular between the SCP and the pick-up coil, which reads out the magnetic field generated by the SCP, located at \(|\boldsymbol{\xi}| = d\) -- here \(d\) is the distance between the equilibrium point of the SCP and the centre of mass of the apparatus. The tidal force that causes the displacement is
\begin{equation}\label{eq:matter_force}
    F_i = \frac{\alphaM M}{\Lambda} \ddot\varphi_{ij}\xi^j.
\end{equation}
We can estimate the magnitude of this force in a typical set-up with \(M=10^{-5}\,\mathrm{g}\) and \(d = \sqrt3\,\mathrm{cm}\) for an SCP whose equilibrium point is at \((x_0,y_0,z_0) = (6\,\mathrm{cm},6\,\mathrm{cm},6\,\mathrm{cm})\) in a \(L=10\,\mathrm{cm}\) shield (see \cref{tab:params}), as
\begin{align}\label{eq:matter_estimate}
    \sqrt{\left<F^2\right>} = \frac{\alphaM M\mfp d \sqrt{\rho_\mathrm{DM}}}{\sqrt3 \Lambda} \approx 2.1\times10^{-10} \alphaM \left(\frac{\mfp}{10^{-12}\mathrm{eV}}\right) \mathrm{eV}^2,
\end{align}
where the \(\left<\cdot\right>\) stands for an average over time and polarisations and we set \(\Lambda=M_\mathrm{Pl}\) with \(M_\mathrm{Pl}\) the reduced Planck mass.\footnote{We assume that the shield dominates the mass budget of the apparatus, which thus determines its centre of mass.} Note that obtaining a non-vanishing tidal force requires the trap to be placed away from the centre of mass of the shield, so that \(d\neq0\). Additionally, it is worth mentioning that while the force \cref{eq:matter_estimate} grows with the distance \(d\), the SCP flux quadrupolar gradient that is read by the SQUID drops according to the distance between the SCP and the pick-up loop as \(1/d_{\rm pick-up}^5\). 

\subsection{Light coupling}
\label{sec:light}

We now focus on the effect of the coupling with light. In this section we derive an estimate of the induced magnetic field at the trap and the resulting driving force on the SCP, \cref{eq:light_estimate} below.

The effect of the coupling between the dark graviton and photons can be schematically represented by the interaction Lagrangian
\begin{equation}
    \mathcal{L}_{\mathrm{int},A} = J^\mu_{\rm{eff}} A_\mu.
\end{equation}
The effective current $J^\mu_{\rm{eff}}$ can be derived from the equations of motion for the field $A_\mu$ including the interaction term \cref{eq:Llight}. The result is:
\begin{align}
\label{eq:Jeff}
    J^\mu_{\rm eff}=\frac{2\alphaL}{\Lambda}\partial_\lambda \left[ \varphi^{\mu\kappa} F_\kappa{}^\lambda + \varphi^{\kappa\lambda} F^{\mu}{}_\kappa\right]
\end{align}
The effective current depends on both the electric and magnetic field. However, a magnetic levitation experiment is performed inside a magnetic shielding, whose effect is to cancel the tangential electric field at the shield surfaces. If the Compton wavelength of the dark graviton is much larger than the characteristic size of the shield, which is the case for this experiment, the electric field does not vary inside the shield itself. The electric field being vanishing at the walls of the shield, it will be vanishing (or more correctly, will be parametrically suppressed) inside the shield too. We can then neglect the electric field contribution to the current. Moreover, neglecting $\varphi_{0\mu}$ thanks to the gradient hierarchy further simplifies the expression for the effective current, leading to $J_{\rm eff}^0 \simeq 0$ and
\begin{equation}
\label{eq:Jeff3D2}
    J_{\rm eff}^i \simeq\frac{2\alphaL}{\Lambda}\partial_j \left[ \varphi^{ik} F_k{}^j + \varphi^{kj} F^{i}{}_k\right].
\end{equation}

\begin{figure}[ht]
\begin{center}
\includegraphics[width=1.0\textwidth]{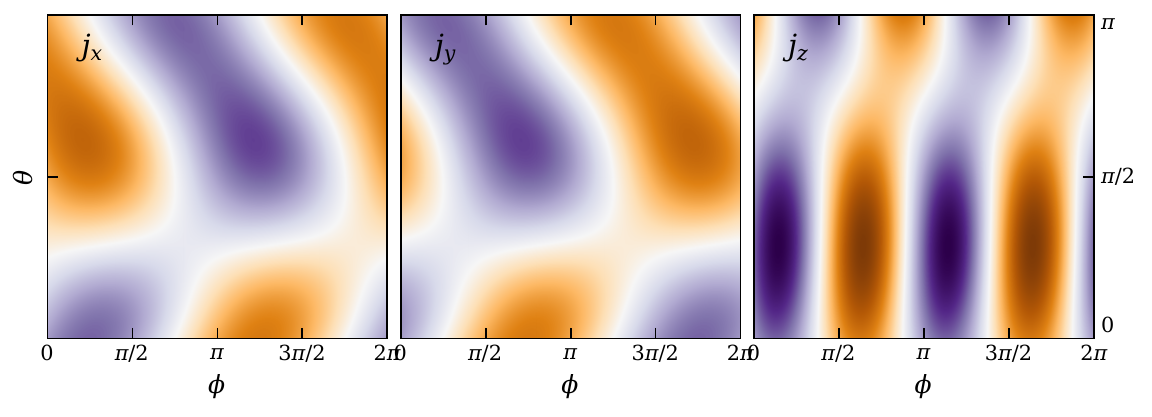}
\end{center}
\caption{Angular currents $j_i$ of \cref{eq:littlej} as a function of the angles $\theta$ and $\phi$ (arbitrary units, shared colour scale across panels), which define the relative orientation of the dark matter basis with respect to the proper detector frame, assuming equipartition among the five dark graviton polarisations. In a realistic case of a superposition of many waves each will have its own polarisation structure and orientation.}
\label{fig:jeffvsthetavsphi}
\end{figure}

Given that the dark graviton is non-relativistic, its dominant effect is to modify the Ampère-Maxwell law as
\begin{align}
    \nabla \times \Bdm-\partial_t\mathbf{E}_\mathrm{DM}=\Jeff.
\end{align}
While the effective current can source, in principle, both an electric and a magnetic field, we once again neglect the electric field inside the cavity because the shield is a perfect conductor. Hence, the magnetic field generated by the effective current satisfies
\begin{align}
\label{eq:Bequation}
    \nabla \times \Bdm \simeq \Jeff.
\end{align}
The magnetic field induced by the dark graviton, \(\Bdm\), generates the Lorentz force responsible for the motion of the SCP. The explicit form of this current is obtained by using \cref{eq:dmansatz} into \cref{eq:Jeff3D2}, from which we find, in the proper detector frame:
\begin{align}
\label{eq:Jeffxyz}
    \Jeff &=-\frac{\sqrt2 \alphaL \varphi_0}{\Lambda}
    \begin{pmatrix}
    j_x (\partial_yB_y-\partial_z B_z) \\
    j_y (\partial_zB_z-\partial_x B_x) \\
    j_z (\partial_xB_x-\partial_y B_y)
\end{pmatrix} \cos{(\mfp t+\zeta)},
\end{align}
where
\begin{align}
\label{eq:littlej}
    j_x \coloneqq C_1 \cos\phi + C_2 \sin\phi, \quad j_y \coloneqq C_2 \cos\phi - C_1 \sin\phi, \quad j_z \coloneqq C_3 \cos 2\phi + C_4 \sin 2\phi,
\end{align}
whose angular dependence is shown in \cref{fig:jeffvsthetavsphi}, and
\begin{alignat}{2}
    C_1 &\coloneqq \varepsilon_L \cos\theta - \varepsilon_\times \sin\theta , 
    &\quad C_2 &\coloneqq \sin\theta \cos\theta \left(\sqrt{3}\,\varepsilon_S - \varepsilon_+\right) + \varepsilon_R \cos 2\theta , \nonumber\\
    C_3 &\coloneqq \varepsilon_L \sin\theta + \varepsilon_\times \cos\theta , 
    &\quad C_4 &\coloneqq \frac{1}{4}\left[\varepsilon_+(3+\cos 2\theta) + 2\varepsilon_R \sin 2\theta + 2\sqrt{3}\,\varepsilon_S \sin^2\theta\right],
\end{alignat}
These expressions have been derived under the assumption that the effective current is sourced by the background magnetic field \cref{eq:Bi}, that is, the trapping field for the SCP. The non-diagonal terms in $b_{ij}$ have been neglected.\footnote{Notice that non-diagonal terms in the magnetic field are a second order effect: indeed, the presence of the dark graviton sources gradient terms in the total magnetic field, which then generates a second order induced magnetic field. This effect is however negligible~\cite{Higgins:2023gwq}.}

We explicitly derive the magnetic field generated by the dark graviton in \cref{app:dmfield}; in particular, we find that, for a trap located at \((x_0,y_0,z_0) = (6\,\mathrm{cm},6\,\mathrm{cm},6\,\mathrm{cm})\) in a \(L=10\,\mathrm{cm}\) shield (as in the matter case), this field is given by
\begin{align}\label{eq:bdm_final}
    \Bdm(\mathbf{r}_0) &\simeq -{\cal A}_B
    \begin{pmatrix}
    1.3j_z-2.4j_y \\
    1.3j_z-2.4j_x \\
    j_x+j_y
\end{pmatrix} \cos{(\mfp t+\zeta)},
\end{align}
where
\begin{equation}
    {\cal A}_B = 1.6\alphaL\,\frac{\sqrt{\rho_\mathrm{DM}} \,b_0\left(R^2+h^2\right)^{5/2}}{\mfp\Lambda L^4}.
\end{equation}

The time-dependent magnetic field will produce a force on the SCP that is the sum of the trapping force and the dark-graviton-driven force
\begin{align}\label{eq:light_force}
    \mathbf{F}_\mathrm{tot}=\mathbf{F}+\mathbf{F}_{\rm DM},
\end{align}
where $F_{\rm DM}\propto \cos{\mfp t}$, resulting in an oscillating motion of the SCP. The strength of the dark-graviton-induced force, for a set-up in which \(h=R=1\,\mathrm{cm}\), \(L=10\,\mathrm{cm}\) can be estimated to be
\begin{equation}\label{eq:light_estimate}
    \sqrt{\left<F^2\right>} \simeq 1.5 \times10^{-14}\alphaL \left(\frac{\mfp}{10^{-12}\mathrm{eV}}\right)^{-1}\mathrm{eV}^2.
\end{equation}

From this discussion we can infer that, because of the dependence of the light force on the geometry of the system, by changing some of the characteristics of the experimental apparatus, such as placing the trap in the centre of the shield, we can make the force \cref{eq:light_estimate} vanish, leaving only the tidal force.

In this section we have hitherto assumed that the light coupling \(\alphaL\) is parametrically larger than the matter coupling. However, strictly speaking, for a non-zero \(\alphaM\) the effective current \cref{eq:Jeff} depends on the difference \((\alphaL-\alphaM)\). We can see this directly by applying our definition of the (matter) proper detector frame. The metric in this frame is built from the \(\tilde{g}_{\mu\nu}\) of \cref{eq:metric_re}, and matter fields experience a locally flat metric (i.e.\ the interaction term \cref{eq:Lmatter} vanishes locally, and the tidal force arises from the matter-driven curvature). On the other hand, the metric governing the electromagnetic sector, in the same matter-defined proper detector frame, receives a contribution $\propto -\alphaM \varphi_{\mu\nu}$, which modifies the interaction \cref{eq:Llight} to \((\alphaL-\alphaM)\varphi_{\mu\nu}T^{\mu\nu}_\mathrm{light}\). In the universal coupling limit a proper detector frame is such for all fields and, as it should, the non-universal interaction vanishes, leaving only the universal interaction with \(\alphaM=\alpha\), which has been absorbed into \(\tilde{g}_{\mu\nu}\). Hence, in this limit, the effective current vanishes at this perturbative order, leaving the tidal force \cref{eq:matter_force} as the dominant effect on the detector.

A notable difference between the dark graviton, the axion/axion-like and dark photon cases is the different frequency dependence of the induced forces. Indeed, for the axion/axion-like and dark photon fields the forces on the SCP are found to be constant in frequency or growing as \(\mfp = 2\pi f\), respectively. In the dark graviton case the force \cref{eq:light_estimate} is \emph{inversely} proportional to the frequency, which means that it is the strongest at the lowest frequencies that we can reliably probe. This has profound implications for detection, as we discuss in the next section.

\section{Sensitivity}
\label{sec:sensitivity}

\subsection{Noise model}

In order to project the expected sensitivity from this experimental architecture, we compare the expected signals from the forces \cref{eq:matter_estimate} and \cref{eq:light_estimate} to the instrumental noise. Following closely the treatment of~\cite{Higgins:2023gwq} (see also~\cite{Hofer:2022chf,Beckey:2023shi}), we write the three noise components relevant for our frequency range as the thermal noise \(S^\mathrm{th}_{FF} = 4M \gamma T\), the imprecision noise \(S^\mathrm{imp}_{FF} = S_{\phi\phi}\, \eta^{-2}|\chi(\omega)|^{-2}\) and the back-action noise \(S^\mathrm{ba}_{FF} = \eta^2 S_{JJ}\) -- here \(S_{\phi\phi}\) is the SQUID flux noise and \(S_{JJ}\) the SQUID current noise. Putting it all together we obtain 
\begin{align}
    S_{FF}^\mathrm{tot} = S^\mathrm{th}_{FF} + S^\mathrm{imp}_{FF} + S^\mathrm{ba}_{FF} = 4M\gamma T + \frac{\kappa}{\tilde\eta^2 \left|\chi(\omega)\right|^2} + \kappa\tilde\eta^2,
\end{align}
where \(\gamma\) and \(T\) are the dissipation rate and temperature of the system and \(\kappa = \sqrt{S_{\phi\phi} S_{JJ}}\) is the SQUID's energy resolution~\cite{Voss:1981uncertainty}. The parameter \(\tilde\eta \coloneqq \eta \sqrt[\leftroot{-1}\uproot{2}4]{S_{JJ}/S_{\phi\phi}}\) is a coupling that we assume can be tuned to match either of two different regimes: a resonant regime in which the coupling saturates \(\tilde\eta^2 \geq 4M\gamma T/\kappa\), where the imprecision (or shot) noise dominates and the sensitivity is maximised at \(\omega=\omega_0\); a broadband regime in which the coupling saturates \(\tilde\eta^2 \leq M\omega_0^2\), where the back-action noise dominates and the sensitivity is maximised at \(\omega\ll\omega_0\).\footnote{Notice that we are ignoring the fact that the imprecision and back-action noises at low frequencies are inversely proportional to \(f\) assuming that the signal is modulated/up-converted to a higher-frequency prior to readout~\cite{Paik:1986mr,Cinquegrana:1993zg}.} Lastly, the mechanical susceptibility is defined as
\begin{equation}
    \chi(\omega) \coloneqq \frac{1}{M\left(\omega_0^2-\omega^2-i\gamma\omega\right)},
\end{equation}
with the angular frequency defined as \(\omega = 2\pi f\) so that the resonant frequency is \(\omega_0 = 2\pi f_0\).

The sensitivity to the dark graviton can be derived by setting a target signal-to-noise (SNR) ratio, which we customarily choose to be \(\mathrm{SNR}=3\). For integration times \(t_\mathrm{int}\) shorter than the DM coherence time \(t_\mathrm{coh} \coloneqq 4\pi / \mfp v_\mathrm{DM}^2 \simeq 10^6 / f\) the dark graviton signal is coherent and the SNR is given by
\begin{equation}\label{eq:snrCoh}
    \mathrm{SNR} = \frac{F_\mathrm{DM}^2}{2S_{FF}^\mathrm{tot}} \, t_\mathrm{int}, \quad\mathrm{for}~~t_\mathrm{int} \leq t_\mathrm{coh}.
\end{equation}
However, because of the stochastic nature of the dark graviton wave superposition -- recall that the local DM configuration comprises a superposition of waves, each with a slightly different velocity, which can be regarded as coherent only within a coherence time -- integration times longer than the coherence time will in effect observe causally disconnected realisations of the DM field, for which the SNR will add in quadrature:
\begin{equation}\label{eq:snrStoc}
    \mathrm{SNR} = \frac{F_\mathrm{DM}^2}{2S_{FF}^\mathrm{tot}} \, \sqrt{t_\mathrm{int} t_\mathrm{coh}}, \quad\mathrm{for}~~t_\mathrm{int} > t_\mathrm{coh}.
\end{equation}

\subsection{Projections}

\begin{figure}[tbhp]
\begin{center}
\includegraphics[width=1.0\textwidth]{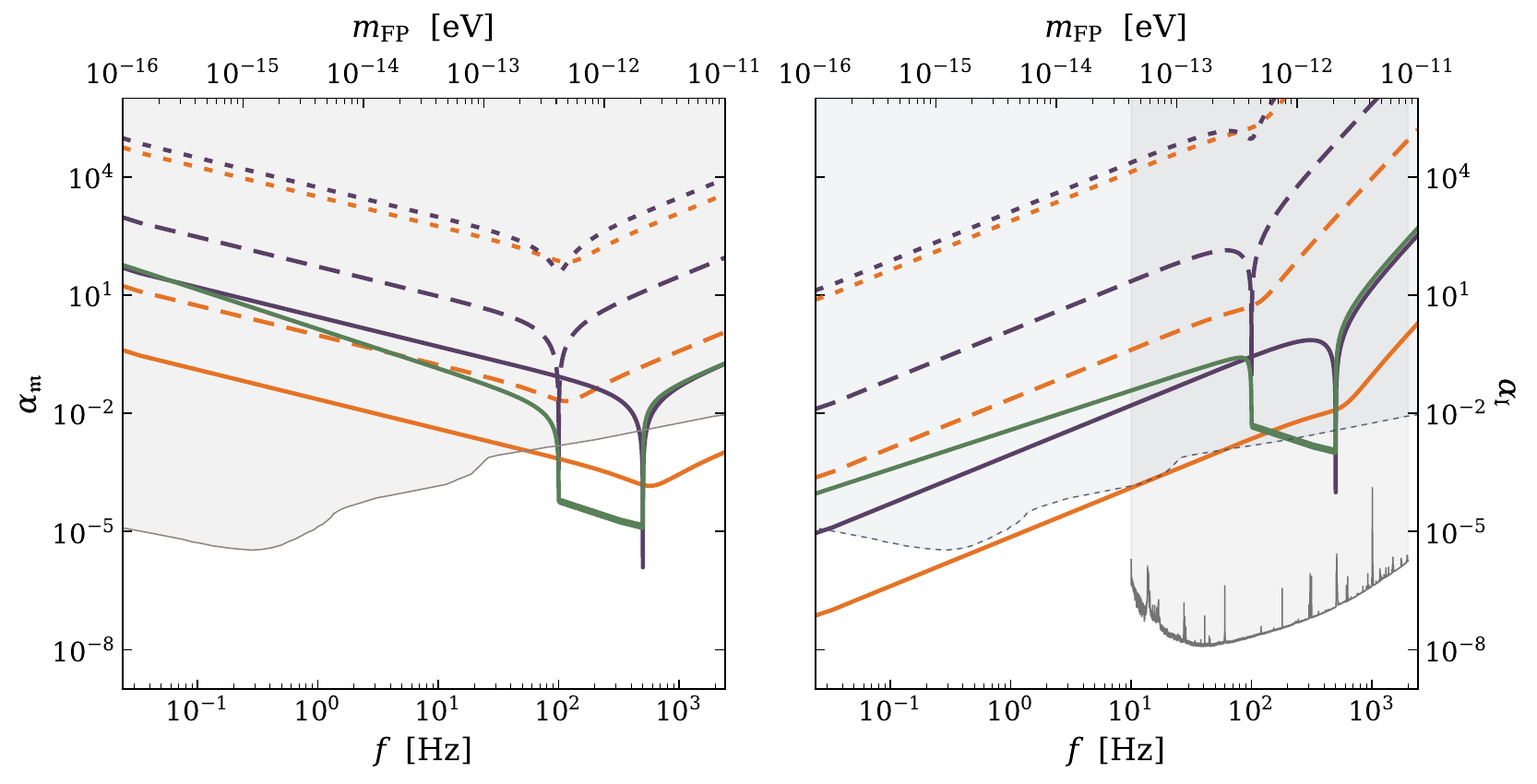}
\end{center}
\caption{Projected sensitivities to the dark graviton coupling to matter \(\alphaM\) (left panel) and light \(\alphaL\) (right panel) as a function of frequency \(f\). The dotted lines refer to the ``baseline'' set-up, the dashed lines are for the ``improved'' set-up and the solid lines are for the ``future'' set-up (see main text). Dark purple lines refer to the resonant regime, dark orange lines to the broadband case and dark green are for a \(1\,\mathrm{yr}\) resonant scan. Existing fifth-force constraints~\cite{Murata:2014nra,Cembranos:2017vgi} are shown in the shaded grey region; the recent LVK limits obtained with the LPSD method are also shown in darker grey~\cite{LIGOScientific:2025ttj}. All these limits apply differently to matter coupling and light coupling, see the main text.}
\label{fig:sensitivity}
\end{figure}

In \cref{fig:sensitivity}, we show the projected sensitivities to the dark graviton coupling to matter \(\alphaM\) (left panel) and light \(\alphaL\) (right panel) as a function of frequency \(f\). We show the sensitivities for three sets of parameters, where in all cases the dark purple lines refer to the resonant regime and the dark orange to the broadband one. The dotted lines refer to the ``baseline'' set-up, which are the same values we have used for the force estimates \cref{eq:matter_estimate} and \cref{eq:light_estimate}; these parameters correspond to the ``existing'' set-up of~\cite{Higgins:2023gwq} and are representative of what can already be built with current technology. The dashed lines in \cref{fig:sensitivity} are for the ``improved'' set-up, which, as discussed in detail in~\cite{Higgins:2023gwq}, could be realistically achieved in the near future. In all cases we assume an integration time of \(1\,\mathrm{yr}\).

\begin{table}[htbp]
\centering
\caption{Numerical parameters adopted for the sensitivity projections.}
\label{tab:params}
\begin{tabular}{@{}lccc@{}}
\toprule
\textbf{Parameter} & \textbf{Baseline} & \textbf{Improved} & \textbf{Future} \\
\midrule
Cavity size $L$              & \SI{10}{\centi\metre}              & \SI{1}{\metre}              & \SI{3}{\metre}              \\[2pt]
Coil radius $R$              & \SI{1}{\centi\metre}               & \SI{10}{\centi\metre}       & \SI{30}{\centi\metre}       \\[2pt]
Coil half-separation $h$          & \SI{1}{\centi\metre}               & \SI{10}{\centi\metre}       & \SI{30}{\centi\metre}       \\[2pt]
SCP density $\rho$           & \SI{10}{\gram\per\centi\metre\cubed} & \SI{0.1}{\gram\per\centi\metre\cubed} & \SI{0.1}{\gram\per\centi\metre\cubed} \\[2pt]
SCP mass $M$                 & \SI{1e-5}{\gram}                   & \SI{1}{\gram}               & \SI{1}{\kilo\gram}          \\[2pt]
Tidal distance $d$               & \SI{1.7}{\centi\metre}            & \SI{17}{\centi\metre}      & \SI{52}{\centi\metre}       \\[2pt]
SCP critical $\mathcal{B}$   & \SI{3.7}{\milli\tesla}             & \SI{80}{\milli\tesla}       & \SI{4}{\tesla}              \\[2pt]
Field gradient $b_0$         & \SI{58}{\tesla\per\metre}         & \SI{5.8}{\tesla\per\metre}  & \SI{29}{\tesla\per\metre}   \\[2pt]
Dissipation rate $\gamma/2\pi$ & \SI{1e-5}{\hertz}              & \SI{1e-8}{\hertz}           & \SI{1e-8}{\hertz}           \\[2pt]
Resonant frequency $f_0$     & \SI{100}{\hertz}                    & \SI{100}{\hertz}            & \SI{500}{\hertz}            \\[2pt]
SQUID resolution $\kappa$    & \num{5}                            & \num{5}                     & \num{1}                     \\
\midrule
Temperature $T$              & \multicolumn{3}{c}{\SI{10}{\milli\kelvin}}                                               \\[2pt]
Integration time $t_{\mathrm{int}}$ & \multicolumn{3}{c}{\SI{1}{\yr}}                                             \\
\bottomrule
\end{tabular}
\end{table}

Finally, the solid lines are for an optimistic ``future'' set-up with a much heavier SCP that relies on the possibility of building larger, \(30\,\mathrm{cm}\)-radius coils within a \(3\,\mathrm{m}\) magnetic shield, while achieving standard-quantum-limit readout \(\kappa=1\). In this futuristic set-up the field gradients would be in the tens of~\(\mathrm{T}/\mathrm{m}\); a SCPs with critical magnetic fields of \(4\,\mathrm{T}\) capable of sustaining such gradients could be realised, for instance, by coating the SCP with a thin film of TiN as proposed in~\cite{Carney:2024zzk}. With this set of parameters we can also devise an optimised scan strategy that runs over resonant frequencies in such a way that, for each frequency bin, the integration time corresponds to the coherence time for that bin, as proposed in~\cite{Higgins:2023gwq}. The solid dark green line in \cref{fig:sensitivity} shows the result of such scan between \(100\,\mathrm{Hz} \leq f \leq 500\,\mathrm{Hz}\).

In order to contextualise the detection power of the levitated SCP set-up, in \cref{fig:sensitivity} we plot the fifth-force limits, obtained from~\cite{Murata:2014nra,Cembranos:2017vgi}, as a shaded light-grey region (see also~\cite{Sereno:2006mw}), whereas the recent LVK limits appear as an overlaid fainter grey region.\footnote{See also~\cite{Delgado:2026hnx} for an optimised implementation that promises to further improve these limits.} Notice that LVK bounds do not directly apply to the matter coupling because they come from light propagation effects~\cite{Gue:2026wqd}; the leading-order pure matter effects will be suppressed by \(v_\mathrm{DM}^2\), pushing the LVK constraints up by a factor of about \(10^6\) and well above the fifth-force limits.\footnote{In the (matter) proper detector frame, the strain due to the tidal force on the mirrors of an interferometer is given by \(h_\mathrm{tidal} \propto \alphaM D^{ij}\varphi_{ij}\), where \(D^{ij}\) describes the detector response~\cite{Armaleo:2020efr}. However, photons do not propagate on the matter metric but on a spatial metric proportional to \((\alphaL-\alphaM)\varphi_{ij}\). Therefore, light propagation will also contribute a strain that is \(h_\mathrm{propagation} \propto (\alphaL-\alphaM)D^{ij}\varphi_{ij}\). Putting them together we see that \(\alphaM\) from the mirror displacement is exactly cancelled by the change in photon propagation.} On the other hand, fifth-force bounds are generally obtained assuming that the coupling between the dark graviton and standard matter is universal (and proportional to the inertial mass), so that for the light coupling alone they are expected to be rescaled by the electromagnetic-to-inertial charge ratio, which determines the relative importance of electromagnetic corrections to baryon masses, and which is of order \(10^{-3}\)~\cite{Damour:2010rp}, leaving \(\alphaL\) much more unconstrained -- this is why, for the light coupling, we choose a dashed charcoal line to report these limits.\footnote{Here we use the estimates for the charge ratio expected for a dilaton field~\cite{Damour:2010rp} only as an order-of-magnitude estimate. We expect that the different index structure of the coupling in the dark graviton case would change this by \(\mathcal{O}(1)\) factors.} In the universal case, when \(\alphaM=\alphaL=\alpha\) the dominant light-coupling force vanishes and both fifth-force and LVK constraints are present, thus limiting the coupling \(\alpha\) to values that are below what the LSD can reach.

\subsection{Discussion}

From the sensitivity projections displayed in \cref{fig:sensitivity} we observe that existing fifth-force constraints already limit the matter coupling \(\alphaM\) to values that are below what the LSD can reach, except in the range \(f\gtrsim50\,\mathrm{Hz}\). In that frequency range a sensitivity to couplings as low as \(\alphaM \simeq 10^{-5}\) could be achieved; this is especially interesting because, if the coupling to light is parametrically suppressed, the LSD, albeit in a set-up that requires significant technological advances, would be able to test yet unexplored regions in the \(\alphaM\) parameter space. 

The levitated SCP set-up is also promising in terms of testing the coupling to light -- particularly if the coupling to matter is parametrically suppressed. In this case, a large region of parameter space opens up for the levitated SCP to probe, especially for \(f\lesssim10\,\mathrm{Hz}\), making the levitated SCP a compelling candidate as a dedicated detector for the dark-graviton--light coupling.

In \cref{fig:sensitivity} we continue our projected sensitivities into the dHz region, a region that remains unexplored in current and planned gravitational-wave detectors. These projections assume that it is possible to maintain the same level of noise control for lower frequencies, which is far from obvious because eliminating the \(1/f\) behaviour of the readout noise becomes harder as we move to lower frequencies, as well as vibrational and other technical noise sources (such as imperfections in the SCP surface or seismic noise) become more relevant~\cite{Carney:2019cio}. In practice, in order to take into account low-frequency noise -- primarily seismic noise -- we can introduce an effective noise term parameterised by a power law:
\begin{align}
    S_{FF}^{\mathrm{seis}}(f) \propto S_{FF}^{\mathrm{tot}}(f_\mathrm{s})\left(\frac{f_\mathrm{s}}{f}\right)^{2\beta},
\end{align}
where \(f_\mathrm{s}\) is the frequency at which seismic motion becomes comparable to the intrinsic force noise and \(\beta>0\), where we can take \(\beta=2\) as a representative value~\cite{Timberlake:2019swe}. Focussing on the light coupling only, from \cref{eq:snrCoh} (or \cref{eq:snrStoc}) we find that, excluding seismic noise, the minimum coupling that we can detect scales as 
\begin{align}
    \alpha_{\mathfrak l}(f) \propto \sqrt{S_{FF}^{\mathrm{tot}}(f)}\,f.
\end{align}
Therefore, in the seismic-dominated regime \(f \ll f_\mathrm{s}\), where \(S_{FF}^{\mathrm{seis}}\gg S_{FF}^{\mathrm{tot}}\), the bound is rescaled according to
\begin{align}
    \alpha_{\mathfrak l}(f) \propto \sqrt{S_{FF}^{\mathrm{seis}}(f)}\,f \propto f^{1-\beta}.
\end{align}
Hence, seismic noise enforces a low-frequency turnover (or floor, depending on the slope \(\beta\)) in the sensitivity to the light coupling at \(f \approx f_\mathrm{s}\), below which the sensitivity projections degrade as \(f^{1-\beta}\).

Standard vibrational isolation techniques are expected to keep seismic noise subdominant above \(f_s \sim 1\,\mathrm{Hz}\)~\cite{Carney:2019cio}, although this remains an open engineering challenge, especially considering our `future' set-up requirements~\cite{Hofer:2022chf,Wiens:2021rsi}. Pushing the turnover frequency \(f_\mathrm{s}\) towards \(0.1\,\mathrm{Hz}\) would require significant active isolation -- for instance~\cite{Ubhi:2022tuw} shows that, once tilt is measured and digitally decoupled, active stabilisation of vibrational noise can reach the \(0.1\,\mathrm{Hz}\) range in translation. In order to push even further to \(f_\mathrm{s} \sim 0.01\,\mathrm{Hz}\), it will most likely be necessary to rely on differential measurements, such that common sources of noise could be eliminated by comparing set-ups with different materials -- which also could in principle enhance the detection reach by the number of detectors in the case of fully correlated readout~\cite{Giovannetti:2011chh,Windchime:2022whs} -- or by going to space, where seismic noise is absent, although other low-frequency technical noise sources become relevant~\cite{Kalia:2024eml}.

Despite the challenges in reaching this range of dark graviton masses, the low-frequency region is especially interesting because the light force experienced by the SCP grows as the dark graviton mass decreases (unlike the axion/axion-like and dark photon cases). Moreover, existing fifth-force constraints become increasingly less stringent as we move below \(1\,\mathrm{Hz}\) and the sensitivity of laser interferometers rapidly worsens at such frequencies. This motivates the exploration of this region as one of the promising frequency ranges in which levitated sensors could, alongside atom interferometers~\cite{Blas:2024kps}, be the most sensitive probes of the dark graviton.

\section{Conclusions and outlook}
\label{sec:conclusion}

In this paper we have discussed the possibility of using a magnetically levitated superconductor as a sensor for dark graviton dark matter, focusing on the frequency range \(2.4\times10^{-2}\,\mathrm{Hz}\lesssim f \lesssim2.4\times10^{3}\,\mathrm{Hz}\), corresponding to dark graviton masses of about \(10^{-16}\,\mathrm{eV} \lesssim \mfp \lesssim 10^{-11}\,\mathrm{eV}\). Building on the existing proposals for this type of technology we have derived the two physically distinct types of forces that the dark graviton exerts on the SCP: a matter-coupling force, see \cref{eq:matter_estimate}, that, in a similar fashion as a (slow and massive) continuous gravitational wave, produces a relative displacement between the SCP and the pick-up loop; a light-coupling force, see \cref{eq:light_estimate}, that perturbs the trapping magnetic fields and therefore causes a superdiamagnetic response of the SCP, which is driven away from its equilibrium position.

In order to determine the projected sensitivities in this set-up we have considered the three primary sources of noise, thermal, imprecision and back-action noise, in both broadband and resonant coupling regimes. Our results in \cref{fig:sensitivity} show that, while current technology has a limited sensitivity reach compared to fifth-force experiments in the same frequency range, in the future this technique could rival fifth-force experiments across over four decades in frequency, albeit in complementary ranges for the matter and light couplings. Moreover, this technology can isolate \(\alphaM\) and test it independently from \(\alphaL\), thereby providing a new way to probe the universality of the dark graviton interaction with the Standard Model.

If frequencies well below \(1\,\mathrm{Hz}\) can be reached, by virtue of the inverse-mass scaling of the light force \cref{eq:light_force}, magnetically levitated SCPs could be the most sensitive probe of the coupling between the dark graviton and light, a coupling that most fifth-force experiments do not directly test. One more notable feature of the coupling to light is that the force it produces increases with decreasing dark graviton mass, which sets it apart from axions and dark photons. However, in case of a detection, a more detailed analysis of the spectral features of the signal alongside the possibility of probing different directions (and therefore directly the polarisation content) with multiple pick-up loops in orthogonal directions would be needed to clearly distinguish different dark matter models~\cite{Amaral:2024rbj}.

Moving forward, there are several extensions to the exploratory results we have presented here that we believe are worth considering. First of all, the same experimental concept can be adapted to higher frequencies, up to MHz and beyond, where a flux-tunable microwave resonator is used to read off the changes in the magnetic flux at such frequencies~\cite{Carney:2024zzk}. This is especially relevant for the coupling between the dark graviton and matter, which in most aspects mimics the effect of a slow and massive continuous gravitational wave, and which could be probed for high frequencies that are beyond the reach of current laser interferometers~\cite{LIGOScientific:2025ttj}. High frequencies at comparable sensitivities could be also reached with a multi-layered stack of dielectric discs, levitated by means of optical lasers~\cite{Arvanitaki:2012cn,Aggarwal:2020umq} and other mechanical sensors such as bulk acoustic wave resonators, magnetic Weber bars and others~\cite{Aggarwal:2025noe}.

A search for the dark graviton with magnetically levitated sensors could already be performed by repurposing existing data from the `POLONAISE pathfinder' experiment, in which a permanent magnet is levitated into a superconducting shield~\cite{Amaral:2024rbj}, originally built to detect small-scale gravity~\cite{Fuchs:2023ajk}  and later recast as a detector for \(U(1)_{B-L}\) gauge boson ULDM.\footnote{The `pathfinder' name is ours: we have chosen it because~\cite{Amaral:2024rbj} does not name their experiment but indicate that POLONAISE is its fully fledged version.} Similarly to the \(U(1)_{B-L}\) gauge boson, the dark graviton would produce a relative displacement between the magnet and the shield; the trap field will also be perturbed by the light coupling in a similar fashion as in our set-up. Moreover, when multiple readouts in orthogonal directions are installed, it would be possible to test the spin of the dark-matter-induced signal and therefore tell apart different dark matter models. While the current data only has a limited frequency reach around \(26.7\,\mathrm{Hz}\), the proposed upgrade, POLONAISE, would be able to probe the \(10\,\mathrm{Hz}\) to \(200\,\mathrm{Hz}\) range, therefore providing a measurement that is independent of laser interferometers~\cite{Amaral:2024rbj}.

Finally, the levitated SCP holds promise to be an exquisite sensor for the coupling between the dark graviton and light, especially if we could break through the low-frequency floor. One exciting possibility in this direction is to levitate a ferromagnet over a superconducting plane, in such a way that the ferromagnet is repelled by an `image' magnetic dipole located below the plane; if the ferromagnet is subject to an ac current, such as those produced by the dark graviton coupling to the magnetic trap, the rotational modes of the ferromagnet can make it to experience libration or precession~\cite{Kalia:2024eml}. In a very ambitious `freefall' scenario in which the ferromagnet is in space, and its motion is detected with an interferometric readout system, low-frequency environmental noises could in principle be avoided. From this perspective, low frequencies, even down to \(10^{-5}\,\mathrm{Hz}\), could be reached. Yet another idea is to use a double-magnet set-up to be able to increase the magnetic field gradients but at the same time keeping the field strength relatively small~\cite{Li:2023wcb}. Taken together, these LSD proposals hold considerable promise to open a new window on the dark graviton.

\appendix

\section{Dark-graviton-induced magnetic field}
\label{app:dmfield}

In this appendix we give the details of the derivation of the dark-graviton-induced magnetic field of \cref{eq:bdm_final}. We begin from the effective current \cref{eq:Jeffxyz}. Solving \cref{eq:Bequation} for $\Bdm$ in order to find the magnetic field generated by the dark graviton is challenging because of the non-trivial background magnetic field configuration. However, we can determine the dependence of the induced magnetic field on the parameters of the experiment. Let us start by considering the trap made of two coils traversed by currents $I$ flowing in opposite directions; the two current loops have a radius $R$, and are separated by a distance $2h$. For each of these loops, the corresponding magnetic field is given by the Biot-Savart law as
\begin{equation}
    \Bloop(\mathbf{r})=\frac{I}{4\pi}\int \frac{\dd\mathbf{l}\times (\mathbf{r}-\mathbf{l})}{|\mathbf{r}-\mathbf{l}|^3},
\end{equation}
where $\mathbf{l}=(R\cos\Theta,R\sin\Theta,0)$ parametrises the loop with radius \(R\) and $\mathbf{r}=(x,y,z)$ represents the distance to the centre of the loop. Combining the contributions from both loops we obtain the trapping magnetic field:
\begin{align}\label{eq:Btrap_exp}
    \Btrap(\mathbf{r})=\Bloop(\mathbf{r}-\mathbf{r}_0-h\hat{z})-\Bloop(\mathbf{r}-\mathbf{r}_0+h\hat{z}),
\end{align}
where $\mathbf{r}_0=(x_0,y_0,z_0)$ is the trap centre.

Let us assume the coils are located inside a rectilinear magnetic shield of dimensions $L_x, L_y, L_z$. The solution of \cref{eq:Bequation} for this geometry can be found by performing a cavity mode decomposition; calling $\mathbf{E}_n$ and $\mathbf{B}_n$ the electric and magnetic cavity modes of the shield, the magnetic field response to the effective current $\bf{J}_{\rm eff}$ is given by
\begin{align}
\label{eq:BDM}
    \Bdm(\mathbf{r}) = \sum_n c_n \frac{\omega_n}{\mfp}\mathbf{B}_n(\mathbf{r})e^{-i \mfp t},
\end{align}
where
\begin{align}
\label{eq:cn}
    c_n \coloneqq \frac{i \mfp}{\omega_n^2-\mfp^2}\frac{\int\dd V \mathbf{E}_n(\mathbf{r})^* \cdot \mathbf{J}_{\rm eff}(t=0)}{\int\dd V|\mathbf{E}_n(\mathbf{r})|^2}.
\end{align}
In the case of a rectilinear cavity described by the three lengths $L_x, L_y, L_z$, the electric and magnetic modes, as well as the frequency modes $\omega_n$, are characterised by three numbers, one for each dimension -- we collectively denote the cavity modes with the shorthand index \(n\) meaning \((l,m,n)\). The frequencies are given by
\begin{equation}
    \omega_n = \omega_{lmn} \coloneqq \pi \sqrt{\frac{l^2}{L_x^2}+\frac{m^2}{L_y^2}+\frac{n^2}{L_z^2}},
\end{equation}
while the modes $\mathbf{E}_{lmn}$ and $\mathbf{B}_{lmn}$ are divided into TE and TM modes (see \cref{sec:cavityModes} for more details).

In order to solve the integrals in \cref{eq:cn} with the current computed in \cref{eq:Jeffxyz} a fully numerical computation is required. Nonetheless, we can obtain an estimate of the parametric behaviour of the induced magnetic field \(\Bdm\); for simplicity, in what follows we choose the shield to be cubic, namely~$L_x,L_y,L_z = L$. Firstly, let us write the trapping magnetic field at the centre of the trap as
\begin{align}
\label{eq:Bcentre}
    \Btrap(\mathbf{r})=b_0\left(\frac{x-x_0}{2},\frac{y-y_0}{2},-z+z_0\right),
\end{align}
where, upon expanding the trap field \cref{eq:Btrap_exp} around the origin of the trap, we find
\begin{equation}
    b_0\coloneqq -3IR^2h/(R^2+h^2)^{5/2}.
\end{equation}
This establishes the relationship between the current \(I\) and the magnetic gradient \(b_0\). Moving forward, we notice that far from the trap at \(|\mathbf{r}-\mathbf{r}_0|\gg(R,h)\) the trap field behaves as a decaying quadrupole:
\begin{equation}
    \left|\Btrap(\mathbf{r})\right| \propto \frac{IR^2h}{\left|\mathbf{r}-\mathbf{r}_0\right|^4}.
\end{equation}
Inserting this expression into the cavity coefficients $c_n$ in \cref{eq:cn} together with the normalisations $\mathbf{E}_{lmn}\propto L^{-3/2}\propto \mathbf{B}_{lmn}$ (see \cref{sec:cavityModes}), and using that in our set-up $\mfp\ll \omega_n$, we find
\begin{equation}
    \left|\Bdm\right| \propto \, \frac{\left(R^2+h^2\right)^{5/2}}{L^4} \frac{\alphaL \varphi_0 b_0}{\Lambda}.
\end{equation}
The exact proportionality constant depends on the specifics of the experimental apparatus and the relative orientation between the dark graviton and the trap as described by \(j_x,j_y,j_z\) -- notice that this coefficient will be zero if we place the trap at the centre of the cavity because of symmetry, but it is sufficient to place the trap off-centre in any of the three directions to obtain a non-zero value. We solve for the overlap integral numerically up to \((l,m,n) = 101\) and, for a trap located at \((x_0,y_0,z_0) = (6\,\mathrm{cm},6\,\mathrm{cm},6\,\mathrm{cm})\) in a \(L=10\,\mathrm{cm}\) shield, we find this factor to be about \(1.1\) (modulo the \(j_i\) factors of \cref{eq:littlej}); putting all this together we arrive at \cref{eq:bdm_final} in the main text.\footnote{Notice that, if the shield is superconducting, the field will be distorted near its boundary in order for the perpendicular magnetic field to vanish~\cite{Kalia:2024eml}; however, in performing the overlap integrals the numerically most relevant contributions come from near the trap itself. We therefore do not expect the distortion at the edge of the shield to contribute significantly.}

\section{Rectilinear cavity modes decomposition}
\label{sec:cavityModes}

For completeness, we collect here the definitions of the cavity modes, indexed by \(l\), \(m\) and \(n\) for a three-dimensional perfectly conducting shield. These modes are split into electric and magnetic modes, each of which is further divided into TE modes, for which $l,m\geq 0,\, n\geq 1$ and $l+m\neq 0$, and TM modes, for which $l,m\geq 1,\, n\geq 0$. The cavity electric modes are:
\begin{align}
\label{eq:ETE}
    E_{\mathrm{TE},\,lmn} &\coloneqq \NTE{lmn}
    \begin{pmatrix}
        \frac{m}{L_y}
        \cos\!\left(\frac{l\pi x}{L_x}\right)
        \sin\!\left(\frac{m\pi y}{L_y}\right)
        \sin\!\left(\frac{n\pi z}{L_z}\right) \\
        -\frac{l}{L_x}
        \sin\!\left(\frac{l\pi x}{L_x}\right)
        \cos\!\left(\frac{m\pi y}{L_y}\right)
        \sin\!\left(\frac{n\pi z}{L_z}\right) \\
        0
    \end{pmatrix}, \\[10pt]
\label{eq:ETM}
    E_{\mathrm{TM},\,lmn} &\coloneqq \NTM{lmn}
    \begin{pmatrix}
        \frac{ln}{L_x L_z}
        \cos\!\left(\frac{l\pi x}{L_x}\right)
        \sin\!\left(\frac{m\pi y}{L_y}\right)
        \sin\!\left(\frac{n\pi z}{L_z}\right) \\
        \frac{mn}{L_y L_z}
        \sin\!\left(\frac{l\pi x}{L_x}\right)
        \cos\!\left(\frac{m\pi y}{L_y}\right)
        \sin\!\left(\frac{n\pi z}{L_z}\right) \\
        -\left(\frac{l^2}{L_x^2}+\frac{m^2}{L_y^2}\right)
        \sin\!\left(\frac{l\pi x}{L_x}\right)
        \sin\!\left(\frac{m\pi y}{L_y}\right)
        \cos\!\left(\frac{n\pi z}{L_z}\right)
    \end{pmatrix},
\end{align}
and the cavity magnetic modes are
\begin{align}
\label{eq:BTE}
    B_{\mathrm{TE},\,lmn} &\coloneqq
    -\frac{\pi i}{\omega_{lmn}}\,\NTE{lmn}
    \begin{pmatrix}
        \frac{ln}{L_x L_z}
        \sin\!\left(\frac{l\pi x}{L_x}\right)
        \cos\!\left(\frac{m\pi y}{L_y}\right)
        \cos\!\left(\frac{n\pi z}{L_z}\right) \\
        \frac{mn}{L_y L_z}
        \cos\!\left(\frac{l\pi x}{L_x}\right)
        \sin\!\left(\frac{m\pi y}{L_y}\right)
        \cos\!\left(\frac{n\pi z}{L_z}\right) \\
        -\left(\frac{l^2}{L_x^2}+\frac{m^2}{L_y^2}\right)
        \cos\!\left(\frac{l\pi x}{L_x}\right)
        \cos\!\left(\frac{m\pi y}{L_y}\right)
        \sin\!\left(\frac{n\pi z}{L_z}\right)
    \end{pmatrix}, \\[10pt]
\label{eq:BTM}
    B_{\mathrm{TM},\,lmn} &\coloneqq
    \frac{i\,\omega_{lmn}}{\pi}\,\NTM{lmn}
    \begin{pmatrix}
        \frac{m}{L_y}
        \sin\!\left(\frac{l\pi x}{L_x}\right)
        \cos\!\left(\frac{m\pi y}{L_y}\right)
        \cos\!\left(\frac{n\pi z}{L_z}\right) \\
        -\frac{l}{L_x}
        \cos\!\left(\frac{l\pi x}{L_x}\right)
        \sin\!\left(\frac{m\pi y}{L_y}\right)
        \cos\!\left(\frac{n\pi z}{L_z}\right) \\
        0
    \end{pmatrix}.
\end{align}
The normalisations are:
\begin{align}
    \NTE{lmn} &\coloneqq
    \sqrt{\frac{2^{3-\delta_{l0}-\delta_{m0}}}
         {\left(\frac{l^2}{L_x^2}+\frac{m^2}{L_y^2}\right)L_x L_y L_z}},
    \label{eq:NTE} \\[6pt]
    \NTM{lmn} &\coloneqq
    \sqrt{\frac{2^{3-\delta_{n0}}}
         {\left(\frac{l^2 n^2}{L_x^2 L_z^2}
               +\frac{m^2 n^2}{L_y^2 L_z^2}
               +\left(\frac{l^2}{L_x^2}+\frac{m^2}{L_y^2}\right)^{\!2}
          \right)L_x L_y L_z}}.
    \label{eq:NTM}
\end{align}

\section*{Acknowledgements}
We are indebted to Dorian Amaral and Jordan Gué for extensive and constructive feedback on this work. PCMD is supported by the Czech Science Foundation (GAČR) project PreCOG (Grant No.\ 24-10780S). VD and FU acknowledge support from the European Structural and Investment Funds and the Czech Ministry of Education, Youth and Sports (project No. FORTE--CZ.02.01.01/00/22\_008/0004632). This article is based upon work from the COST Action COSMIC WISPers CA21106, supported by COST (European Cooperation in Science and Technology).

\bibliographystyle{JHEP}
\bibliography{main}

\end{document}